\documentclass[showpacs, pra,onecolumn,preprintnumbers ,amsmath, amssymb, superscriptaddress, aps]{revtex4-2}
\usepackage{color}
\usepackage{amsmath,amssymb}
\usepackage{pifont}
\usepackage{amssymb}  
\usepackage{bbold}
\usepackage{float}
\usepackage{subfloat}

\usepackage[caption=false]{subfig}
\usepackage{tikz}
\usepackage{makecell}
\usepackage{subfig}
\usepackage{pifont}   
\usepackage{graphicx} 
\graphicspath{{Figures/}}
\usepackage{dcolumn}  
\usepackage{bm}       
\usepackage{multirow} 
\usepackage{placeins}
\usepackage[colorlinks]{hyperref}
\usepackage{mathtools}
\usepackage{appendix}

\begin{document}
	
	\title{Tunneling in  rippled graphene superlattice with spin dependence and a mass term }
	\date{\today}
	\author{Jaouad El-hassouny}
	\affiliation{Laboratory of Nanostructures and Advanced Materials, Mechanics and Thermofluids, Faculty of Sciences and Techniques, Hassan II University, Mohammedia, Morocco}
	
	\author{Ahmed Jellal}
	\email{a.jellal@ucd.ac.ma}
	\affiliation{Laboratory of Theoretical Physics, Faculty of Sciences, Choua\"ib Doukkali University, PO Box 20, 24000 El Jadida, Morocco}
	\affiliation{Canadian Quantum  Research Center,
		204-3002 32 Ave Vernon,  BC V1T 2L7,  Canada}
	\author{El Houssine Atmani}
	\affiliation{Laboratory of Nanostructures and Advanced Materials, Mechanics and Thermofluids, Faculty of Sciences and Techniques, Hassan II University, Mohammedia, Morocco}
	
	
	\begin{abstract}
		
		The insertion of the band gap $\Delta$ in the rippled graphene superlattice leads to new outcomes, as demonstrated. The essential thing is the appearance of opposite-spin transmissions, which increase with $\Delta$ and vanish without it. Furthermore, compared to the $\Delta=0$ scenario, the duration of the suppression of the transmission with the same spin is longer, with many peaks.
		The maximum value of transmissions with the same spin declines and remains around unity. Furthermore, for particular energy values, a shift in the behavior of the transmission channels is found.
		As a result, we demonstrate that with $\Delta$, transmission filtering becomes crucial. 
		Finally, as a result of the band gap, distinct variations in total conductance are discovered. 

	\end{abstract}

	\maketitle

	\section{Introduction}

Graphene is a zero-gap semiconductor  \cite{Novoselov,Zhang} with band structure described by a low-energy linear dispersion relation, like massless Dirac-Weyl fermions \cite{dresselhaus1998physical,neto2009electronic}. Such a band structure leads to the exceptionally high conductivity of the charged carriers. Because of its importance in the technological field, finding possible solutions that would manage the mobility and, therefore, control the conductivity is a hot topic in graphene physics. Experimentally, there are different techniques to open a gap in a graphene band structure \cite{Abergel2010} and its maximum value could be 260 meV due to the non-symmetry of the sublattice \cite{Enderlein2010}. 
By manipulating the structure of the graphene-ruthenium interface \cite{Giovannetti2007} or epitaxed graphene on a SiC substrate \cite{Enderlein2010,zhou}, an energy gap can be measured. Moreover, other methods can be used to open and adjust the band gap of graphene, such as diffusion, which could be induced by a ripple created in a sheet of graphene
 \cite{Katsnelson2008,alyobi2019modifying}. 
 Ripples act as potential barriers for charged carriers, leading to their localization \cite{vasic2016}. Additionally, periodically repeated ripples can be created and controlled in suspended graphene, notably by thermal treatment \cite{Bao2009} or by placing the graphene in a specifically prepared substrate. 
 
 Theoretically, alternative studies have been proposed to create an energy gap in graphene systems. For example, the scattering of electrons through a periodically repeated rippled chain (the superlattice)  was investigated in \cite{puldak2020,smotlacha}. This study demonstrated that the graphene superlattice leads to the suppression of electron transmission with one spin orientation on the contrary to the other. 
 Another method investigated the scattering of electrons through the superlattice, considering a chain formed by concave and convex undulations to create a sinusoidal type  graphene sheet \cite{pudlak2015cooperative}. As a result, the spin filtering effect, which is weak in one ripple, becomes important with an increasing number of connected ripples \cite{pudlak2015cooperative}. 
 In our previous work, \cite{jaouad2021}, we demonstrated that adding a band gap to repplied graphene reduces transmission channels with spin-up/-down but increases those with spin opposite. We detected resonances in reflection channels with the same spin, indicating that backscattering with a spin-up/-down is not null in ripple. It is found that the spin filter is affected by some critical band gap values, resulting in a reduction of the channels. Furthermore, we demonstrated that the band gap, as opposed to the null gap \cite{pudlak2015cooperative}, affects total conductance.


We extend our previous work \cite{jaouad2021} to a rippled graphene superlattice, introducing a band gap and investigating quantum tunneling. We derive the energy spectrum and use the matrix transfer to analytically obtain the full transmission and reflection chancels. We show that adding a band energy acts by decreasing the transmissions with spin-up/-down but increasing those with spin opposite. We also observe that electron transmission is strongly suppressed through the superlattice for some energy values. Furthermore, we show that the total conductance gets shifted by the band gap, in contrast to the case of gap null \cite{puldak2020,smotlacha}. We conclude  that the presence of a band gap can be used as a key tool to control the transport properties of our superlattice.

	The following is how the present paper is organized.
	In section \ref{Tmod}, we develop our theoretical model and solve the Dirac equation for the $j$-th cell to obtain the energy spectrum for each region. In section \ref{Tmod1}, using the boundary conditions at interfaces of $2N$ points and the matrix transfer method, we compute all the transmission and reflection channels in addition to the corresponding conductance. 
	We numerically analyze the transmission channels and conductance under various conditions to highlight their basic features in section \ref{BB}. Finally, we provide a summary of our results.

		\section{Theoretical model}\label{Tmod}
	
\begin{figure}[H]
		\centerline{\includegraphics[scale=0.4]{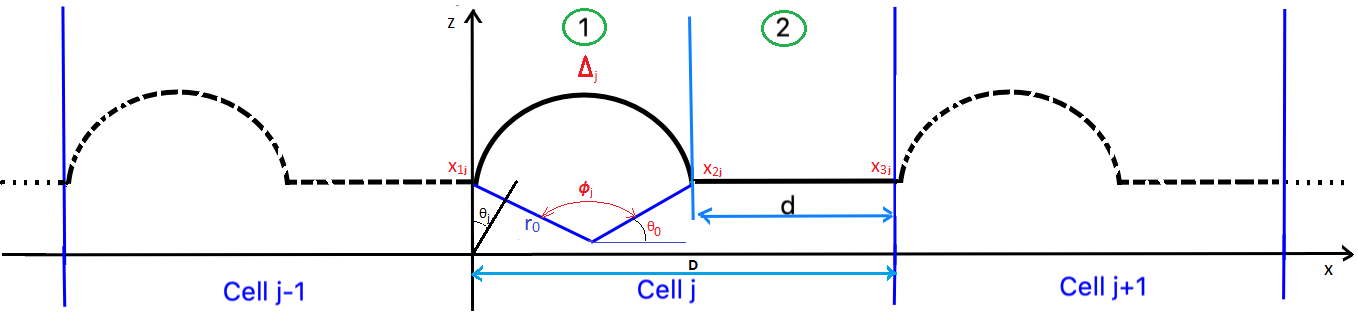}}
		\caption{(color online) The profile of graphene is made up of $ N $ unit cells, each of which is an arc (ripple) of radius $r_0$ with a mass term $\Delta$ connected to the planar part  of length $d$. The specific parts are linked at the connection points ($1$ to $2N$). }
		\label{fig11}
	\end{figure}

	We examine an indefinitely large corrugated graphene, as seen in Fig.~\ref{fig11}.
	It consists of $ N $ elementary cells designated by $j\ (j=0, \cdots, N-1)$, each of which is made up of a juxtaposition of two regions: one arc of a circle with radius $r_0$ and angle $\phi_j$ involving a mass term $\Delta_j$, and a flat graphene with length $d$.	
%
	The $j$-th elementary cell has one internal junction at $x_2j=2r_0\sin\phi_j/2+j D$ and two extreme junctions at $x_1j=j D, x_3j=(j+1) D)$. The width of one cell is $D = d + 2r_0\sin\phi_j/2$.
	 We assume ${W} \gg {L},$ ($W$ and $L$ being the width along the $y$- and length along the $x$-axes of our system, respectively) and that the electrons are infused into the superlattice at $k_y=0$ if edge impacts are ignored. 
	 According to Fig.~\ref{fig11}, for $j$-th elementary cell the gap $\Delta_{j}$ and the angle of ripple $\theta_{j}$ are given by
	\begin{equation}
		\Delta_{j}, \theta_{j}=\left\{\begin{array}{lll}
		\Delta_, \theta	& \text { if } \quad & j D<x<2r_{0}\sin{\phi_{j}/2}+j D \\
			 0 & \text { if } \quad & 2r_{0}\sin{\phi_{j}/2}+j D <x<(1+j) D
		\end{array}\right.
	\end{equation}
	Our graphene superlattice composed of $ N $ unit cells,
which are placed between the input and output regions.
In region 1 of j-th elementary cell, as depicted in Fig.~\ref{fig11}, has
the Hamiltonian spin dependent \cite{jaouad2021,ando2000spin}
	\begin{equation}\label{eqq8}
	\mathcal{H}_1^{j}
	= 
\begin{pmatrix}
	0 & 0 & -\gamma\partial_{y}-\Delta_{j}-\mathrm{i} \frac{\gamma}{r_{0}} \partial_{\theta_{j}} & \mathrm{i}\left(\lambda_{y}+\lambda_{x}\right) \\
	0 & 0 & \mathrm{i}\left(\lambda_{y}-\lambda_{x}\right) &-\gamma\partial_{y}+\Delta_{j}-\mathrm{i} \frac{\gamma}{r_{0}} \partial_{\theta_{j}} \\
	\gamma\partial_{y}-\Delta_{j}-\mathrm{i} \frac{\gamma}{r_{0}} \partial_{\theta_{j}} & -\mathrm{i}\left(\lambda_{y}-\lambda_{x}\right) & 0 & 0 \\
	-\mathrm{i}\left(\lambda_{y}+\lambda_{x}\right) & \gamma\partial_{y}+\Delta_{j}-\mathrm{i} \frac{\gamma}{r_{0}} \partial_{\theta_{j}} & 0 & 0
\end{pmatrix}.
\end{equation}
We get four bands $ E_{j1s}^{s'} $ at normal incidence, $k_{y}=0,$ by solving the eigenvalue equation
\begin{equation}\label{eqq18}
	E_{1js}^{s'}=s'\sqrt{t_{m_{j}}^{2}+\lambda_{y}^{2}}+s\sqrt{\Delta_{j}^{2}+\lambda_{x}^{2}}
\end{equation}
where  $ s,s'=\pm $ and 
$
t_{m_{j}}=\frac{\gamma}{r_0} m_{j}$, $t_{y}=\gamma k_{y}$, $\lambda_{x}=\frac{\gamma}{2 r_0}(1+4 \delta p)$,  $\lambda_{y}=\frac{\delta \gamma^{\prime}}{4 r_0}$ have been defined. As a result, we can calculate the angular momentum as follows: 
\begin{equation}\label{eqq24}
	m_{js}^{s'}=s'\frac{r_{0}}{\gamma} \sqrt{\left(E_{j1s}^{s'}-s \sqrt{\Delta_{j}^2+\lambda_{x}^2} \right)^{2}-\lambda_{y}^{2}} 
\end{equation}
and here
$m_{j}=m_{js}^{s'}$. The eigenspinors then   take the form
\begin{equation}
	\Psi_{1}^{jss'}(\theta_{j}, m_{j})=
	\begin{pmatrix}
		\cos\frac{\theta_{j}}{2}A_{ss'}(m_{j}) -	\sin\frac{\theta_{j}}{2}	B_{ss'}(m_{j})\\
		\sin\frac{\theta_{j}}{2}A_{ss'}(m_{j}) +	\cos\frac{\theta_{j}}{2}	B_{ss'}(m_{j})\\
		\cos\frac{\theta_{j}}{2}-	\sin\frac{\theta_{j}}{2}D_{s}(m_{j})\\
		\sin\frac{\theta_{j}}{2}+	\cos\frac{\theta_{j}}{2}D_{s}(m_{j})
	\end{pmatrix}
	e^{i m_{j} \theta_{j}}
\end{equation}
and the quantities 
\begin{align}
	&A_{jss'}(m_{j})=s'\frac{\lambda_{x} \sqrt{t_{m_{j}}^{2}+\lambda_{y}^{2}}+s \lambda_{y} \sqrt{\Delta_{j}^{2}+\lambda_{x}^{2}}}{\lambda_{x} t_{m_{j}}-\lambda_{y} \Delta_{j}}
	\\
	&	B_{jss'}(m_{j})=- i s'\frac{\Delta_{j} \sqrt{t_{m_{j}}^{2}+\lambda_{y}^{2}}+s t_{m_{j}} \sqrt{\Delta_{j}^{2}+\lambda_{x}^{2}}}{\lambda_{x} t_{m_{j}}-\lambda_{y} \Delta_{j}}\\
	&	C=1
	\\
	&D_{js}(m_{j})=- i\frac{t_{m_{j}}\Delta_{j} +\lambda_{y}\lambda_{x}+s\sqrt{(t_{m_{j}}^{2}+\lambda_{y}^{2})(\Delta_{j}^{2}+\lambda_{x}^{2})}}{\lambda_{x} t_{m_{j}}-\lambda_{y} \Delta_{j}}
\end{align}
have been set.
Finally, in region 1 the wave function can be composed as a superposition of four solutions for a rippled part
\begin{equation}
	\Psi_{1}^{j}(\theta_{j})=	a_{+j} \Psi_{1}^{+}\left(\theta_{j}, m_{j+}^+\right)
	+b_{+j} \Psi_{1}^{+}\left(\theta_{j}, m_{j+}^-\right)+a_{-j} \Psi_{1}^{-}\left(\theta_{j}, m_{j-}^+\right)
	+b_{-j} \Psi_{1}^{-}\left(\theta_{j},m_{j-}^-\right)
\end{equation}
with	$a{_{\pm}}_{j}$ and $ b{_{\pm}}_{j},$ denote the coefficients of the linear combination. It can be expressed as a matrix as 
\begin{equation}
	\Psi_{1}^{j}(\theta_{j})=	M_{1}^{}(\theta_{j})
	\left(e^{\mathrm{i} {m_{j}}\theta_{j}}\right) 
	 D_{1j}
\end{equation}
	such that 
\begin{align}
&	M_{1}^{}(\phi_{}/2)=\Lambda_{}{M}_{1}
\\
&	\Lambda_{}=\begin{pmatrix}
		\cos \frac{\phi_{} }{4} & -\sin  \frac{\phi_{}  }{4} & 0 & 0 \\
		\sin \frac{\phi_{}  }{4} & \cos  \frac{\phi _{} }{4} & 0 & 0 \\
		0 & 0 & \cos \frac{\phi _{} }{4} & -\sin  \frac{\phi _{} }{4} \\
		0 & 0 & \sin  \frac{\phi _{} }{4} & \cos  \frac{\phi _{} }{4}
	\end{pmatrix}
	\\
	& \label{eq26}
	{M}_{1} =\begin{pmatrix}
		A_{+,s'}(m_{+}) & A_{-,s'}(m_{-}) & A_{+,s'}(-m_{+}) & A_{-s'}(-m_{-}) \\
		B_{+,s'}(m_{+}) & B_{-,s'}(m_{-}) & B_{+,s'}(-m_{+}) & B_{-,s'}(-m_{-}) \\
		1 & 1 & 1 & 1 \\
		D_{+}(m_{+}) & D_{-}(m_{-}) & D_{+}(-m_{+}) & D_{-}(-m_{-})
	\end{pmatrix} \\
&
	\left(e^{\mathrm{i} {m_{j}}\theta_{j}}\right) 
	=	\text{diag}\left(e^{\mathrm{i} m_{j+} \theta_{j}}, e^{\mathrm{i} m_{j-} \theta_{j}}, e^{\mathrm{-i} m_{j+} \theta_{j}}, e^{\mathrm{-i} m_{j-} \theta_{j}}
	\right)
\\
&
	D_{1j}=\begin{pmatrix}
		a_{+j} \\
		b_{+j}\\
		a_{-j} \\
		b_{-j}
	\end{pmatrix}.
\end{align}

For flat part, regions 2,  by solving the eigenvalue equation, we can derive the two band energy at $k_y=0$  \cite{jaouad2021}
\begin{equation}
	E_{j2}=\pm \gamma|k_{j}|
\end{equation}
We can also write the relationship $ E_{j2}=E^{s'}_{1js} $ because of energy conservation. 
The linked eigenstates have the following structure: 
\begin{equation}
	\psi_{2}^{j,\tau}(x, k_{j})=\frac{1}{2}
	\begin{pmatrix}
		\text{sign}(k_{j}) \\
		\text{sign}(k_{j}) \	i\tau \\
		1 \\
		i	\tau 
	\end{pmatrix}
	e^{i k_{j}  x}
\end{equation}
where  $ \tau=\pm$, $ k_{j}=k_{x} $ and    $\text{sign}(k)=\pm$ refers to  conductance and valence bands, respectively. As a result
the wave function  
can be written a superposition of four possible solutions for planar graphene
\begin{equation}
	\psi_{2}^{j}(x, k_{j})=\alpha_{j}\psi_{2}^{+}(x, k_{j})+\beta_{j}\psi_{2}^{-}(x, k_{j})+\gamma_{j} \psi_{2}^{+}(x,-k_{j})+\xi_{j} \psi_{2}^{-}(x,-k_{j})
\end{equation}
 with $\alpha_{j}, \beta_{j}$, $\gamma_{j}$ and $\xi_{j}$ indicate the coefficients of the linear combination. It can be mapped as
\begin{equation}
	\psi_{2}^{j}(x,k_{j})=	M_{2}  \left(e^{\mathrm{i} {K_{j}} x}\right)
	D_{2j}
\end{equation}
such that 
\begin{align}
&	M_{2}=\begin{pmatrix}
		1 & 1 & -1 & -1 \\
		i  & -i & -i & i \\
		1 & 1 & 1 & 1 \\
		i & -i & i & -i
	\end{pmatrix}
\\
&
	\left(e^{\mathrm{i} {K_{j}} x}\right)=
	\text{diag}\left(e^{\mathrm{i} k_{j} x}, e^{\mathrm{i} k_{j} x}, e^{\mathrm{-i} k_{j} x}, e^{\mathrm{-i} k_{j} x}\right)
\\
&
	D_{2j}=\begin{pmatrix} 
		\alpha_{j} \\
		\beta_{j}\\
		\gamma_{j} \\
		\xi_{j}
\end{pmatrix}.
	\end{align}

As we will see the above-obtained solutions will be utilized to examine transport processes involving transmission and conductance. 

	\section{Transport properties}\label{Tmod1}
	
	The transfer matrix is calculated using the eigenspinors' continuity at $2N$ connection points. We recall that
	the $j$-th elementary cell has one internal junction at
	\begin{align}
	x_2j=2r_0\sin\phi_j/2+j D	
	\end{align} 
	and two extreme junctions at 
	\begin{align}
		x_1j=j D,\qquad  x_3j=(j+1) D).
	\end{align}
	The width of one cell is 
	\begin{align}
	D = d + 2r_0\sin\phi_j/2.	
	\end{align}
	The wave functions of the adjoining components are equal at these connection locations.
	This fact leads to the $2N$ equations system, which yields 
	\begin{align}
	\label{eq13}
	&	\begin{array}{c}
		M_{2}D_{in}=M_{1}\left(-\frac{\phi}{2}\right) \left(e^{-\mathrm{i} {m} \frac{\phi}{2}}\right)
		D_{1,1 } 
	\end{array}\\
	&
	M_{1}\left(\frac{\phi}{2}\right)
	\left(e^{\mathrm{i} {m} \frac{\phi}{2}}\right) 
	D_{1, 1}=M_{2} \ 
	\left(e^{-\mathrm{i} {K} x_{1} }\right)
	D_{2,1} 
	\\
	&
	\begin{array}{c}
		M_{2}\ 
	\left(	e^{\mathrm{i} {K} x_{2}}\right) D_{2,1}=M_{1}\left(-\frac{\phi}{2}\right) 
\left(	e^{-\mathrm{i} {m} \frac{\phi}{2}}\right) D_{1,2}
	\end{array}
	\\
	&~~~~\vdots\qquad\qquad\qquad \qquad\vdots\qquad\qquad\qquad \qquad\vdots
	\nonumber\\
	&		M_{2}  \left(e^{\mathrm{i} {K} x_{N-1}}\right) D_{2,N-1}=M_{1}\left(-\frac{\phi}{2}\right) 
	  \left(e^{-\mathrm{i} {m} \frac{\phi}{2}}\right) D_{1, N} 
	\\
	&
	M_{1}\left(\frac{\phi}{2}\right) 
	 \left(e^{\mathrm{i} {m} \frac{\phi}{2}}\right) D_{1,N}=M_{2}D_{out}\label{eq17}
\end{align}
where the input $D_{{in}}$ and the output $D_{{out}}$ of the superlattice are given by
\begin{equation}\label{eq22}
	D_{in}=\begin{pmatrix}
		\alpha \\
		\beta\\
		r_{\uparrow}^{\xi} \\
		r_{\downarrow}^{\xi}
	\end{pmatrix},\qquad
	D_{out}=\begin{pmatrix}
		t_{\uparrow}^{\xi} \\
		t_{\downarrow}^{\xi} \\
		0 \\
		0
	\end{pmatrix}.
\end{equation}
We take the parameters $ \alpha=0, \beta=1$ for spin down polarization (${\xi}={\downarrow}$) and $\alpha=1, \beta=0$  for spin up polarization ($\xi=\uparrow$) for displacing the electron flux. 
Based on the preceding arguments, we can construct the following link between $D_{in}$ and $D_{out}$:
	\begin{equation}\label{eq21}
D_{in}=M_{2}^{-1} \Omega  M_{2} D_{out}=\mathcal{M} D_{out}
\end{equation}
and the matrix $\Omega$ takes the form
\begin{equation}
	\Omega=M_{2}\left[C\left(e^{-i {K} d}\right)\right]^{N-1} C M_{2}^{-1}
\end{equation}
%
where
\begin{equation}
	C=M_{2}^{-1}\left[M_{1}\left(-\frac{\phi}{2}\right) 
	 \left(e^{-i {m} \phi}\right) M_{1}^{-1}\left(\frac{\phi}{2}\right)\right] M_{2}=
	\left({C}_{kl}\right) 
\end{equation}
has been defined, with $k,l=1,2,3,4  $.

 Now by setting 
\begin{equation}
	\mathcal{M}=\left[C\left(e^{-i {K} d}\right)\right]^{N-1} C=
	\left(\mathcal{M}_{kl}\right) 
\end{equation}
we write \eqref{eq21} as follows
\begin{equation}\label{eq211}
	D_{in}=\mathcal{M} D_{out}
\end{equation}
After a lengthy and straightforward algebraic calculation, we obtain
	the transmission amplitudes for spin-up and spin-down
	\begin{align}
		&t_{\uparrow}^{\uparrow}=\frac{\mathcal{M}_{22}}{\mathcal{M}_{22}\mathcal{M}_{11}-
			\mathcal{M}_{21}\mathcal{M}_{12}}\\	&
		t_{\downarrow}^{\downarrow}=\frac{\mathcal{M}_{11}}{\mathcal{M}_{22}\mathcal{M}_{11}-\mathcal{M}_{21}\mathcal{M}_{12}}
\\
		&t_{\downarrow}^{\uparrow}=-\frac{\mathcal{M}_{21}}{\mathcal{M}_{22}\mathcal{M}_{11}-
			\mathcal{M}_{21}\mathcal{M}_{12}}\\ &
		t_{\uparrow}^{\downarrow}=-\frac{\mathcal{M}_{12}}{\mathcal{M}_{22}\mathcal{M}_{11}
			-\mathcal{M}_{21}\mathcal{M}_{12}}.
	\end{align}
Because the input and output wave vectors are the same, the corresponding transmission and reflection may be expressed as 
	\begin{align}
		T_{\uparrow}^{\uparrow}=	|t_{\uparrow}^{\uparrow}|^{2}, \qquad& T_{\downarrow}^{\downarrow}= |t_{\downarrow}^{\downarrow}|^{2}\\
			T_{\uparrow}^{\downarrow}= |t_{\uparrow}^{\downarrow}|^{2}, \qquad&
		T_{\downarrow}^{\uparrow}= |t_{\downarrow}^{\uparrow}|^{2}.
\\
		R_{\uparrow}^{\uparrow}=	|r_{\uparrow}^{\uparrow}|^{2}, \qquad& 
		R_{\downarrow}^{\downarrow}= |r_{\downarrow}^{\downarrow}|^{2}\\
		R_{\uparrow}^{\downarrow}= |r_{\uparrow}^{\downarrow}|^{2}, \qquad &
		R_{\downarrow}^{\uparrow}= |r_{\downarrow}^{\uparrow}|^{2}.
	\end{align}
We may easily verify that the probability conservation criterion 
\begin{align}
\sum_{p, q=\xi} \left[ \left(T_{p}^{q}\right)^2 + \left(R_{p}^{q}\right)^2\right]= 1.	
\end{align}

At zero temperature, the Landauer-Buttiker formula \cite{buttiker,landauer,bundesmann} can be used to calculate the conductance of our system. It was demonstrated that the conductance of a single multiplying channel can undoubtedly be summarized into any number of incoming and outgoing channels \cite{horacio1,horacio2,lucas}.
There are various conductances arising from the transmittances between states of definite momentum and spin projection at the contacts \cite{imry}. Then, the total conductance in the linear response system is given by adding up all the transmission channels
	\begin{align}
		G=\frac{e^{2}}{h}\left(T_{\uparrow}^{\uparrow}+T_{\downarrow}^{\downarrow}+T_{\uparrow}^{\downarrow}+T_{\downarrow}^{\uparrow}\right)=\frac{e^{2}}{h} \sum_{p, q=\xi} T_{p}^{q}.
	\end{align}
%
 We now attempt to compute the transmission and conductance in various energy domains numerically after getting closed form formulas. This will help us better understand how different physical parameters affect the transmission and conductance
 of rippled graphene superlattice. 

	\section{Numerical results}\label{BB}
	
\begin{figure}[H]
		\centerline{
			\subfloat[]{
				\includegraphics[scale=0.4]{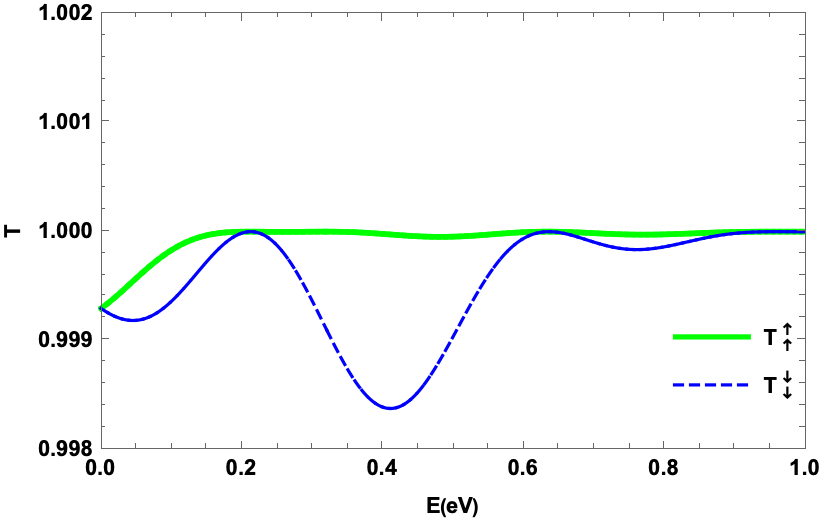}
				\label{subfigurea}}
			{\subfloat[]{
					\includegraphics[scale=0.4]{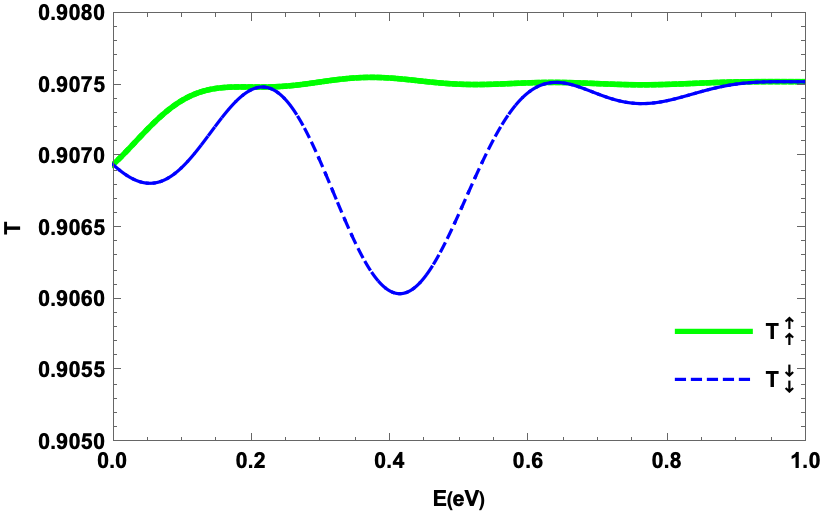}
					\label{subfigureb}}
				{	\subfloat[]{
						\includegraphics[scale=0.4]{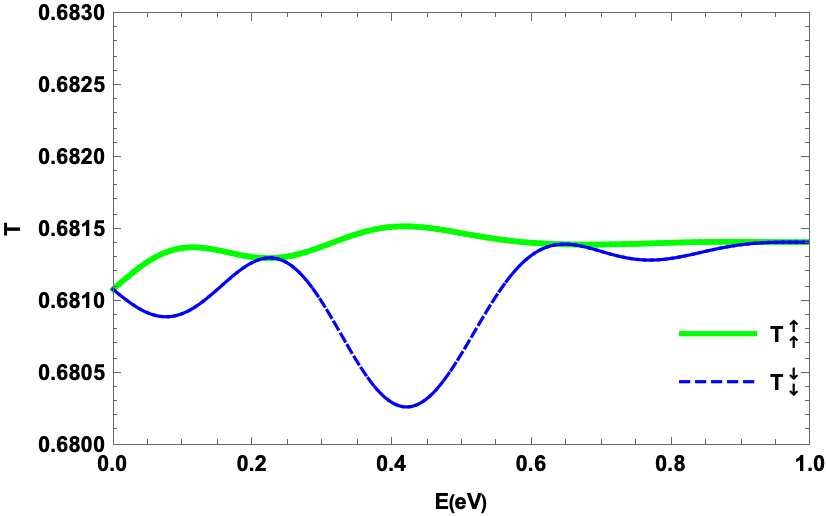}
						\label{subfigurec}}}}}
		\centerline{
			\subfloat[]{
				\includegraphics[scale=0.4]{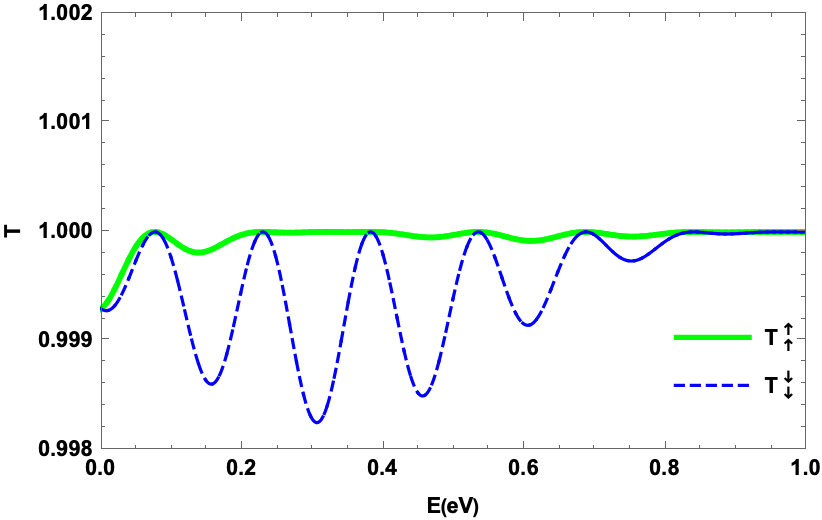}
				\label{subfigured}}{
				\subfloat[]{
					\includegraphics[scale=0.4]{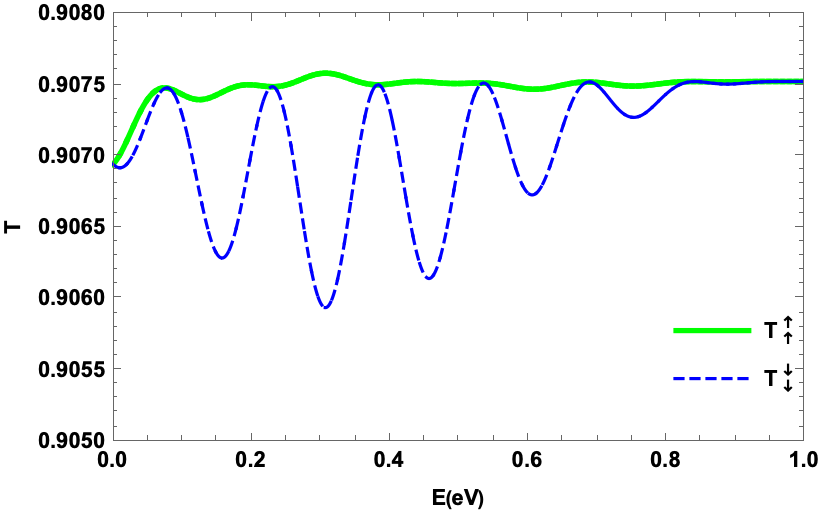}
					\label{subfiguree}}{
					\subfloat[]{
						\includegraphics[scale=0.4]{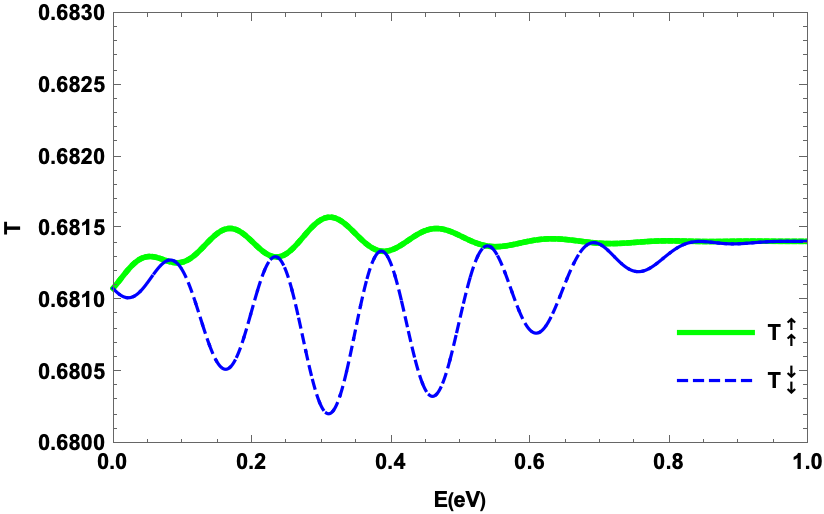}
						\label{subfiguref}}}}}
		\caption{(color online) Transmission $T_{\uparrow}^{\uparrow}$ (green) and $T_{\downarrow}^{\downarrow}$ (blue dashed) 
			at $k_{y}=0$ as a function of energy $E$ for three values of the gap    (a,d): $\Delta=0$,  (b,e): $\Delta=0.05$, (c,f): $\Delta=0.1$. 
			 We chose $\phi=\pi$,  $r_{0}=10\ \text{\AA{}}$,   (top panel): $d=16\ \text{\AA{}}$,  (bottom panel): $d=100\ \text{\AA{}}$ and $N = 2$.}
		\label{fig2}
	\end{figure}

	At $k_{y}=0$, we investigate the tunneling properties of our superlattice using appropriate values for the relevant parameters ($r_0, \Delta, \phi,d,N $).
%
For some values of the gap with $\phi=\pi, r_{0}=10\ \text{\AA{}}, N=2$, $d=16\ \text{\AA{}}$ (up), $d=100\ \text{\AA{}}$ (bottom), Fig.~\ref{fig2}  illustrates the transmissions $T_{\uparrow}^{\uparrow}$ (green) and $T_{\downarrow}^{\downarrow}$ (blue dashed) with the same spin as a function of energy $ E $. As a preliminary consequence, we see that by increasing the gap,
$T_{\uparrow}^{\uparrow}$ and $T_{\downarrow}^{\downarrow}$ decrease.
Now, in the upper panels, we choose the distance between the arcs in the structure $d=16\ \text{\AA{}}$, and for $\Delta = 0$, we see that the $T_{\uparrow}^{\uparrow}$ and $T_{\downarrow}^{\downarrow}$  behavior varies slowly close to the unit, as shown in Fig.~\ref{subfigurea}. 
When we choose $\Delta = 0.05$ in Fig.~\ref{subfigureb}, we see that it decreases $T_{\uparrow}^{\uparrow}$ but has no effect on $T_{\downarrow}^{\downarrow}$. 
According to Fig.~\ref{subfigurec}, when $ \Delta = 0.1 $,
$T_{\uparrow}^{\uparrow}$  decreases to stabilize at a minimum and $T_{\downarrow}^{\downarrow}$  decreases as well, but by oscillating.
The bottom panels are the same as before, with the exception that the distance $d = 100\ \text{\AA{}}$ has been chosen.
As a result, increasing $\Delta$ causes both 
$T_{\uparrow}^{\uparrow}$ and $T_{\downarrow}^{\downarrow}$  to drop, as seen Figs.~(\ref{subfigured}, \ref{subfiguree}, \ref{subfiguref}). 
%
The number of oscillations in the comparing graphs increases as the distance between the arcs in our superlattice grows, while the variation interval remains constant.
The efficiency of the spin filter is negligible for the parameters examined with $\Delta=0$, and the transmission fluctuates about unity, as indicated \cite{smotlacha}. In contrast, the insertion of the gap $\Delta$ has an effect on transmission by lowering its maximum value and increasing filtering efficiency, which was not discovered in the previous study. 
	
\begin{figure}[H]
		\centerline{
			\subfloat[]{
				\includegraphics[scale=0.4]{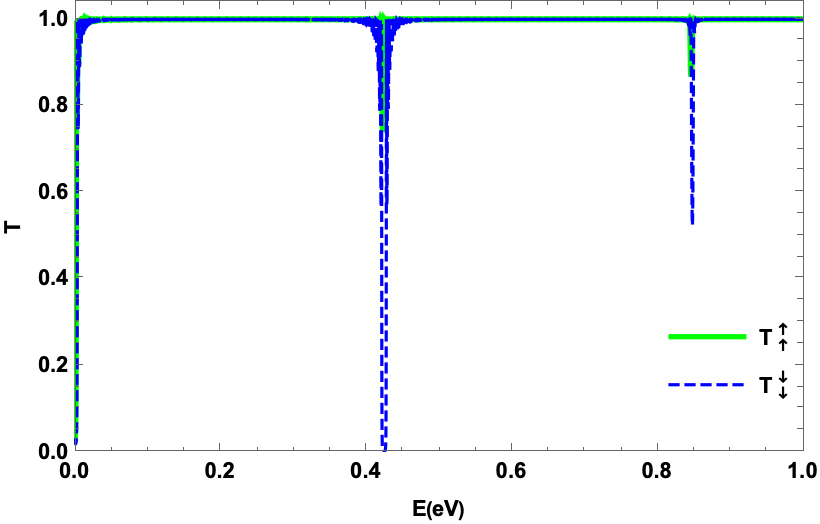}
				\label{subfigure3a}}
			{\subfloat[]{
					\includegraphics[scale=0.4]{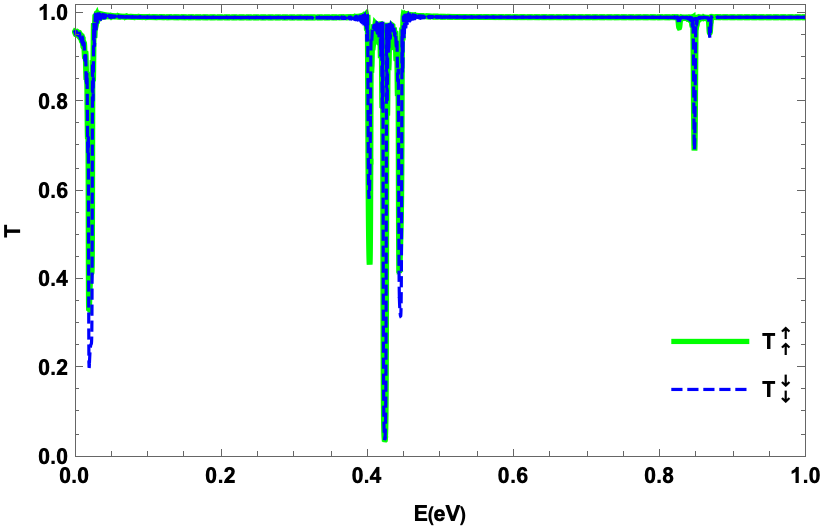}
					\label{subfigure3b}}
				{	\subfloat[]{
						\includegraphics[scale=0.4]{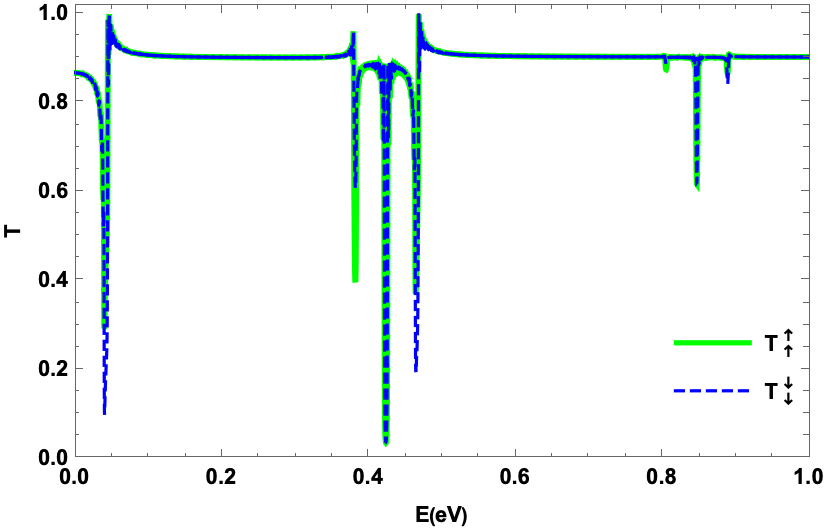}
						\label{subfigure3c}}}}}
		\centerline{
			\subfloat[]{
				\includegraphics[scale=0.4]{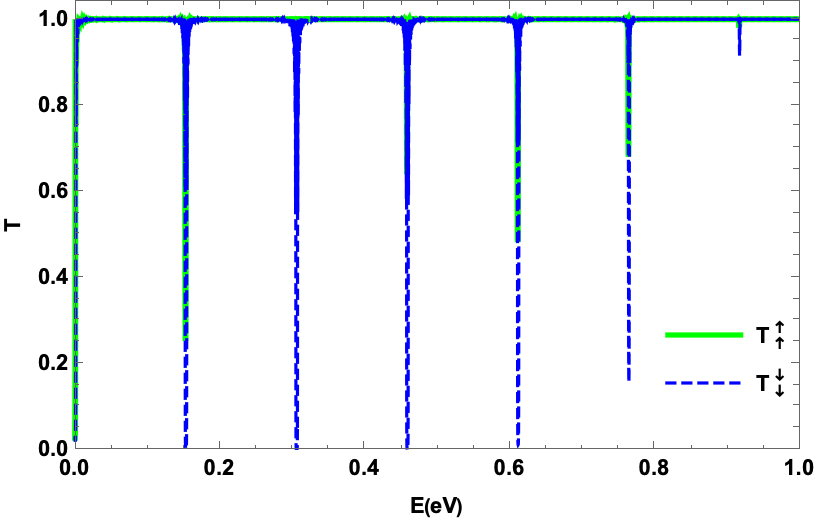}
				\label{subfigure3d}}{
				\subfloat[]{
					\includegraphics[scale=0.4]{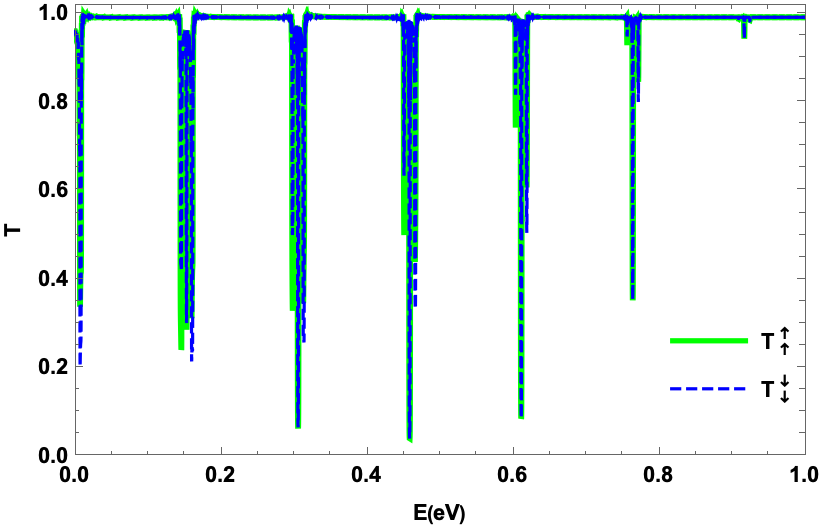}
					\label{subfigure3e}}{
					\subfloat[]{
						\includegraphics[scale=0.4]{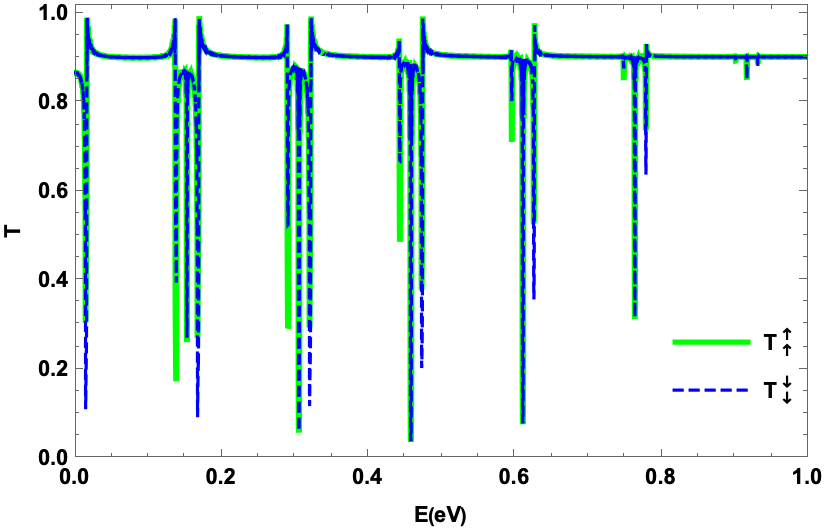}
						\label{subfigure3f}}}}}
		\caption{(color online) 
			The same as in Fig. \ref{fig2}, but now for 
			$N = 200$ cells.}
		\label{fig3}
	\end{figure}

Fig.~\ref{fig3} shows $ T_{\uparrow}^{\uparrow} $ (green) and $T_{\downarrow}^{\downarrow} $ (blue dashed) with same spin  as a function of the energy $E$ but now by increasing the cell number to 
$N=200$.
We notice that the transmissions completely change their behaviors by showing different peaks. When we choose $\Delta = 0$ in Fig.~\ref{subfigure3a}, we see that $T_{\downarrow}^{\downarrow} $
is strongly suppressed in comparison to $ T_{\uparrow}^{\uparrow} $.
The behavior of both transmissions varies and decreases for $\Delta= 0.05$, as shown in Fig.~\ref{subfigureb}. 
$T_{\downarrow}^{\downarrow}$  suppresses almost identically to $ T_{\uparrow}^{\uparrow} $, but with more peaks and a narrower interval between peaks. 
In Fig.~\ref{subfigure3c}, for $\Delta= 0.1$, we see that $T_{\downarrow}^{\downarrow}$  is suppressed nearly as much as $ T_{\uparrow}^{\uparrow} $, but with more peaks. The interval between peaks becomes a bit wider as well as some shifts in the transmissions at the origin are observed.
The number of peaks grows as $ d $ is increased, as shown in 
Fig.~(\ref{subfigure3d},  \ref{subfigure3e}, \ref{subfigure3f}), and the space between the peaks becomes quite small.
Furthermore, notice how the increase in $\Delta$ results in the formation of additional peaks with extremely distinct spacing. 

	\begin{figure}[H]
\centerline{
\subfloat[]{
	\includegraphics[scale=0.42]{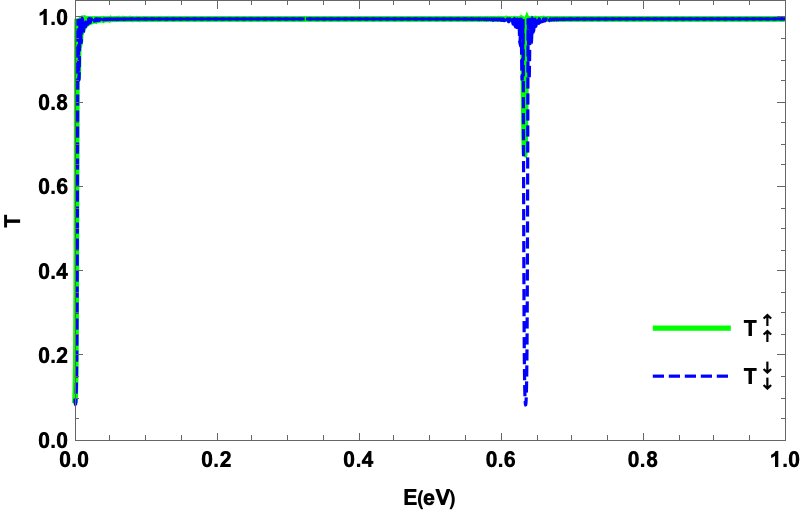}
	\label{subfigure4a}}
{\subfloat[]{
		\includegraphics[scale=0.42]{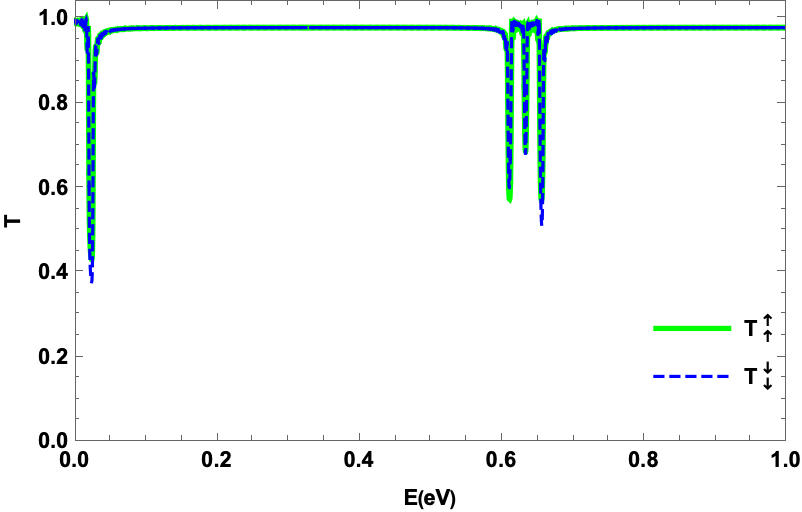}
		\label{subfigure4b}}
	{	\subfloat[]{
			\includegraphics[scale=0.42]{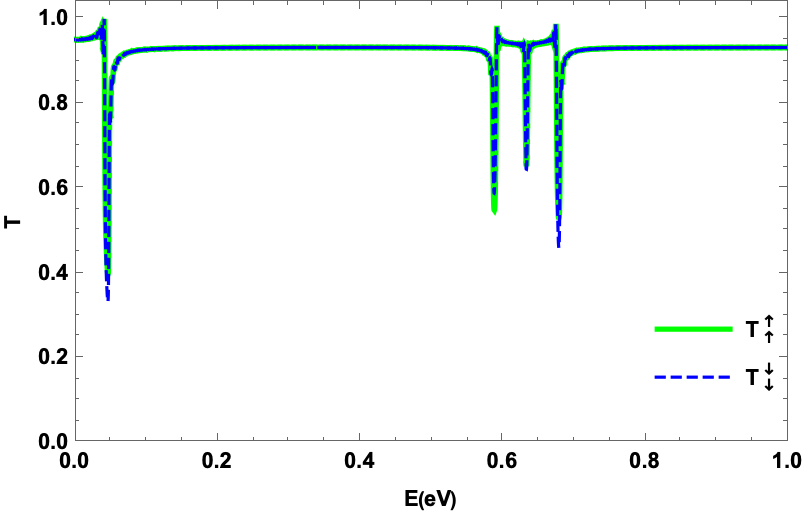}
			\label{subfigure4c}}}}}
\centerline{
\subfloat[]{
	\includegraphics[scale=0.42]{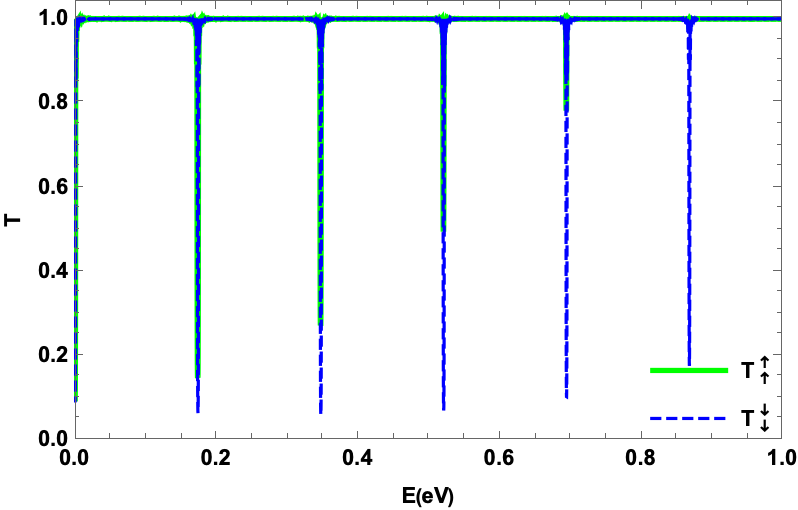}
	\label{subfigure4d}}{
	\subfloat[]{
		\includegraphics[scale=0.42]{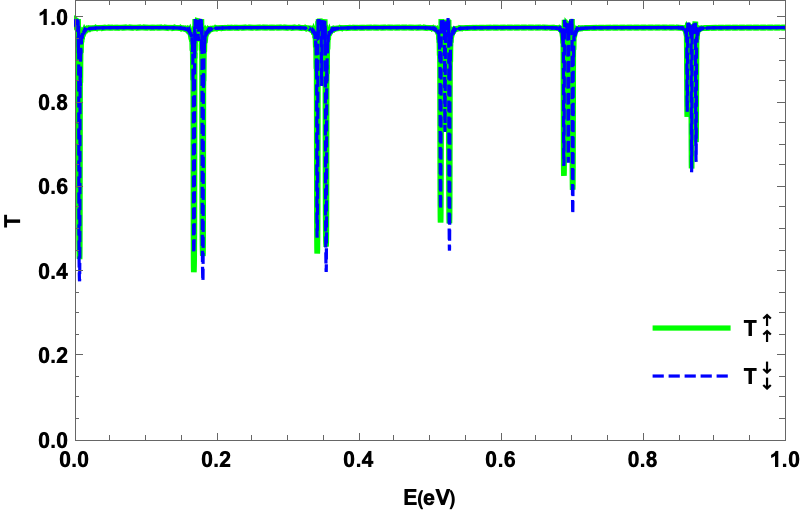}
		\label{subfigure4e}}{
		\subfloat[]{
			\includegraphics[scale=0.42]{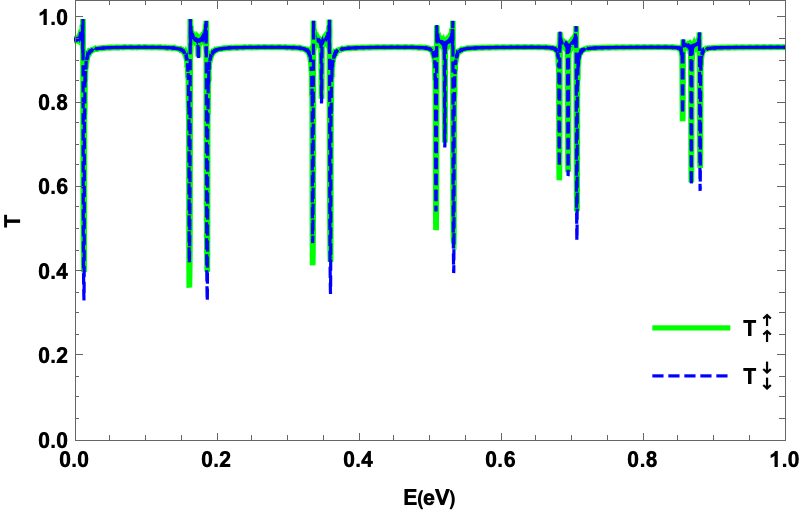}
			\label{subfigure4f}}}}}
\caption{(color online) 
The same as in Fig. \ref{fig3}, but now for $\phi=\pi/2$.}		
		\label{fig4}
\end{figure}

Fig.~\ref{fig4} depicts $ T_{\uparrow}^{\uparrow} $ (green) and $T_{\downarrow}^{\downarrow} $ (blue dashed) with the same spin as a function of energy $E$ for three  values of gap  $\Delta$, with $\phi=\frac{\pi}{2}$, $d=16$\ \text{\AA{}} (top panel), $d=100$\ \text{\AA{}} (bottom panel), $r_{0}=10\ \text{\AA{}}$ and  $ N=200 $. For $\Delta=0$, the suppression of the transmission is reached for two peaks  as clearly seen in Fig.~\ref{subfigure4a}. 
If $\Delta$ increases,  however, Figs.~(\ref{subfigure4b},  \ref{subfigure4c}) 
show the same behavior as before, with the exception that there are numerous peaks with shifts. Furthermore, it is demonstrated that the transmission does not achieve its minimal value as previously stated.  
	We can see that the transmission is canceled at multiple peaks, as illustrated
	in Fig.~\ref{subfigure4d}.
	However, if we increase $\Delta$, as shown in 
	Figs.~(\ref{subfigure4e},  \ref{subfigure4f}), the maximum value of the transmission decreases, new peaks form, but the minimum value does not reach zero, in contrast to the case where $\phi=\pi$ was mentioned above. 
 The angle $\phi=\pi/2$ does not allow the total suppression of transmission but its reduction. Similarly if we take $d=100$\ \text{\AA{}}, except that the interval between the peaks becomes narrow Fig.~(\ref{subfigure4d},  \ref{subfigure4e},  \ref{subfigure4f}). 
 As a result, we can see that when the angle of ripple is reduced, the transmission suppression is not complete, but the effect of $\Delta$ remains the same. We find that $\phi$ can be utilized as a tunable to modify transmission, among other physical parameters.
%


\begin{figure}[H]
	\centerline{
		\subfloat[]{
			\includegraphics[scale=0.42]{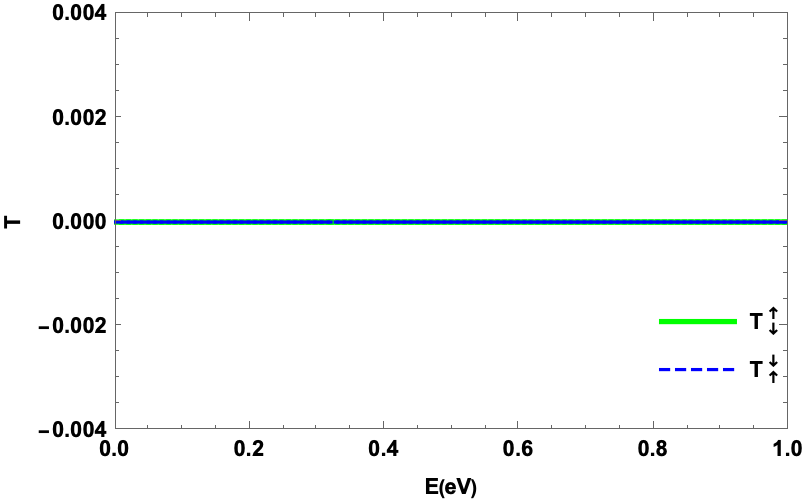}
			\label{subfigure5a}}
		\subfloat[]{
			\includegraphics[scale=0.4]{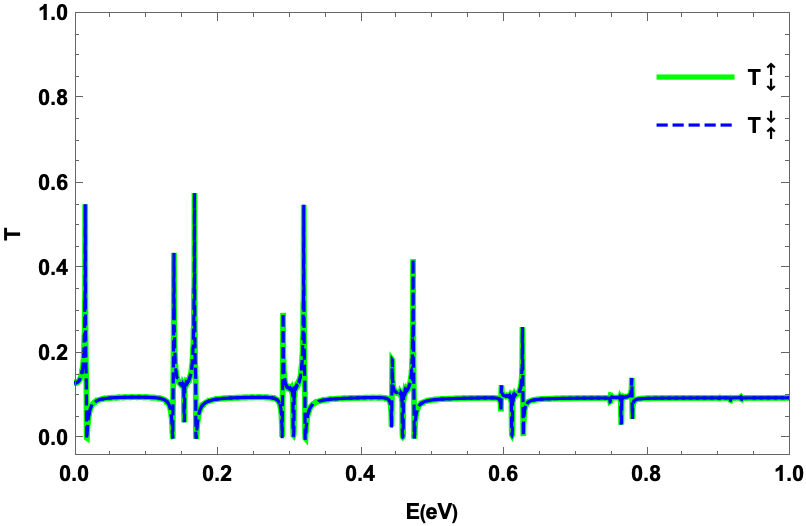}
			\label{subfigure5c}}
		{\subfloat[]{
				\includegraphics[scale=0.4]{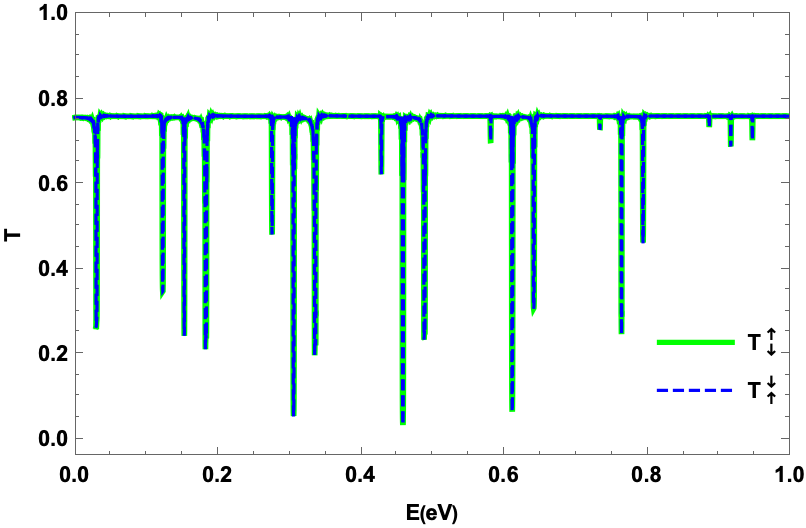}
				\label{subfigure5d}}
	}}
	\caption{(color online) Transmissions of spin inversion $T_{\downarrow}^{\uparrow}$ (green) and $T_{\uparrow}^{\downarrow}$ (blue dashed) 
		at $k_{y}=0$ as a function of the energy $E$ for various values of the gap    (a): $\Delta=0$,  
		(b): $\Delta=0.1$, (c): $\Delta=0.2$  with $\phi=\pi$,  $r_{0}=10\ \text{\AA{}}$, $d=100\ \text{\AA{}}$, $N = 200$.}
	\label{fig5}
\end{figure}

	\begin{figure}[H]
	\centerline{
		\subfloat[]{
			\includegraphics[scale=0.4]{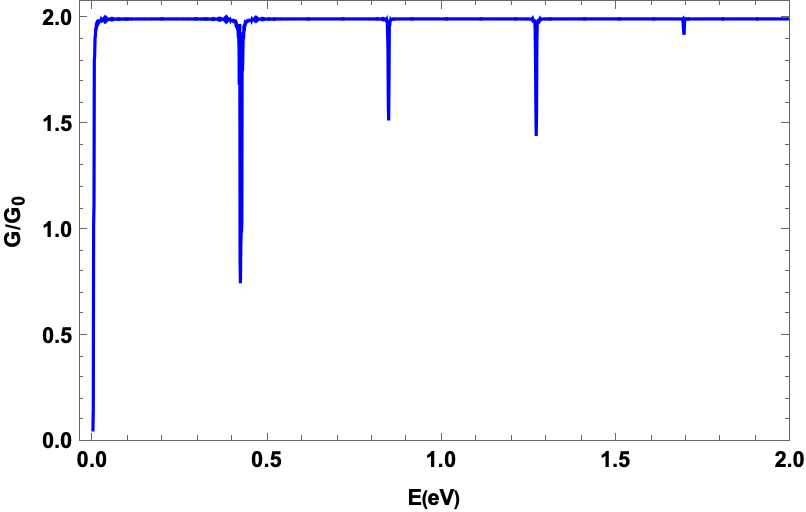}
			\label{subfigure6a}}
		{\subfloat[]{
				\includegraphics[scale=0.4]{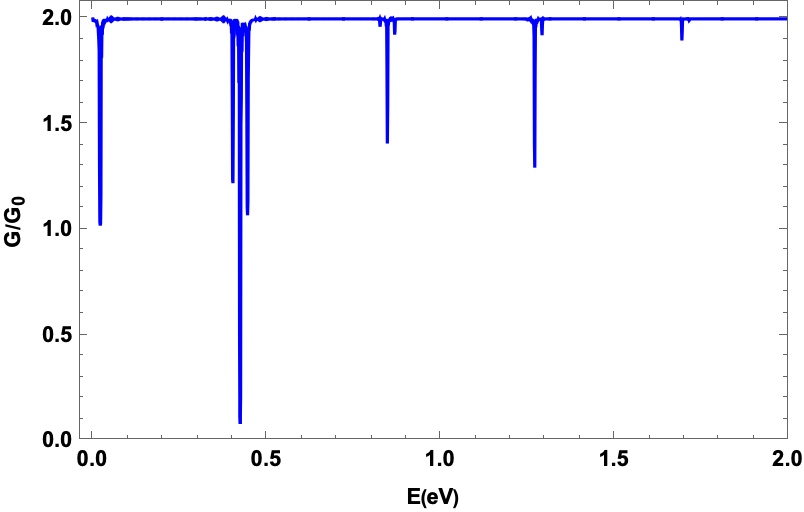}
				\label{subfigure6b}}
			{	\subfloat[]{
					\includegraphics[scale=0.4]{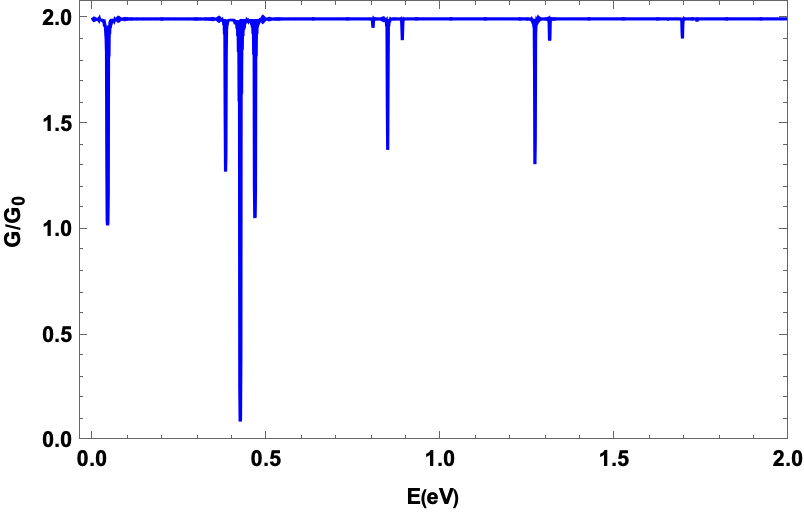}
					\label{subfigure6c}}}}}
	\centerline{
		\subfloat[]{
			\includegraphics[scale=0.4]{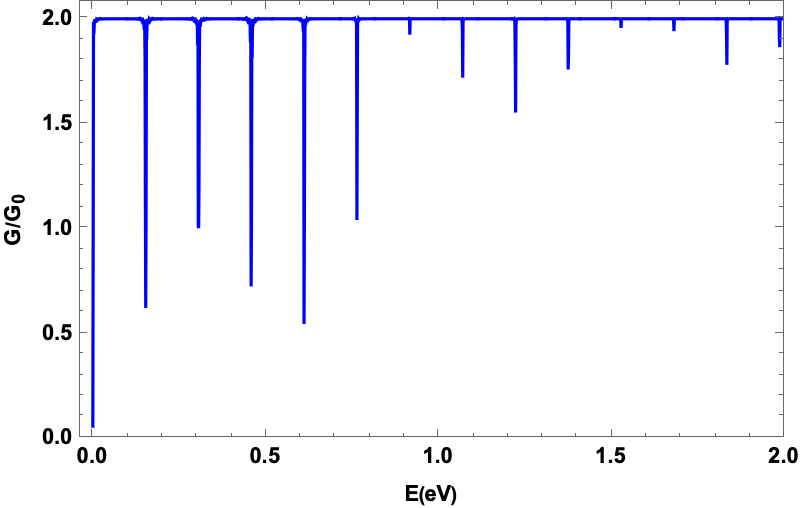}
			\label{subfigure6d}}{
			\subfloat[]{
				\includegraphics[scale=0.4]{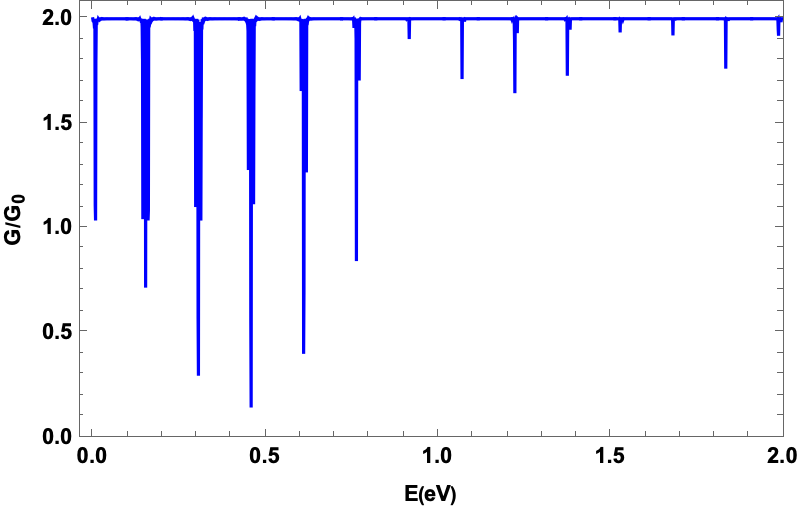}
				\label{subfigure6e}}{
				\subfloat[]{
					\includegraphics[scale=0.4]{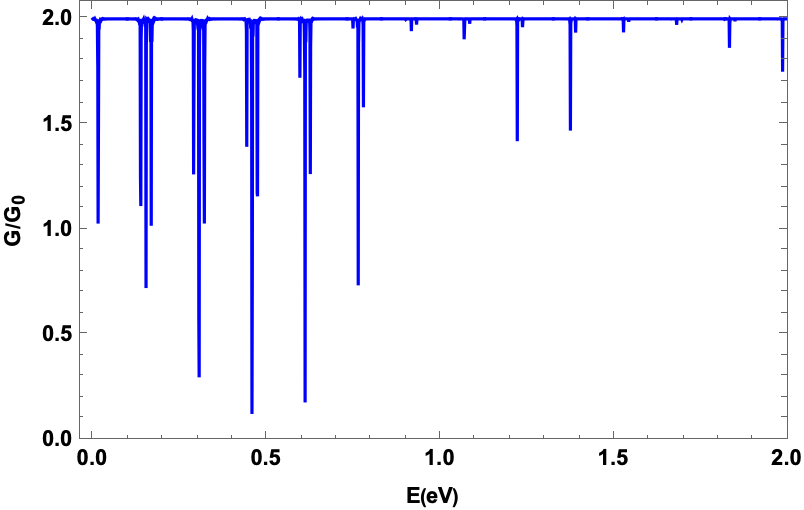}
					\label{subfigure6f}}}}}
	\caption{(color online) 
		Conductance at $k_{y}=0$ as a function of the energy $E$ with the same parameters as Fig. \ref{fig3}.}	
	\label{fig6}
\end{figure}

We show the transmissions of spin inversion 
  $T^{\uparrow}_{\downarrow}$ (green) and $ T^{\downarrow}_{\uparrow}$  (dashed blue) as a function of the energy $E$ for three different gap values  $\Delta$ in Fig.~\ref{fig5} with  $\phi=\pi$, $r_{0}=10 \ \text{\AA{}}$, $d=100 \ \text{\AA{}}$, $N=200$. 
  We see that the two transmissions always behave in the same way, resulting in the relationship
 $T_{\uparrow}^{\downarrow}=T_{\downarrow}^{\uparrow}$. 
 In Fig.~\ref{subfigure5a}, it is evident that the two transmissions are null for $\Delta=0$, as shown in
 \cite{pudlak2015cooperative}. 
 However, for $\Delta\neq0$, it is clear that the transmission increases progressively by increasing the number of peaks as $\Delta$ increases, as shown in Fig. \ref{subfigure5c}, which shows the transmission behavior as peaks up and peaks down. The peaks down reach the zero value, but the peaks up take the maximum values.
%
Surprisingly, for $\Delta=0.2$ eV, the transmission behaves radically differently.
As seen in Fig.~\ref{subfigure5d}, the transmission corresponding to peaks up is inhibited, and the suppression becomes more powerful. 
 As a result, with the existence of $\Delta$, we were able to retrieve the transmission of spin inversion and control and suppress it, which was not possible in the previous study \cite{smotlacha,puldak2020,pudlak2015cooperative}. 
Consequently, we observe that the introduction of $\Delta$ greatly changes the behavior of all transmission channels and  can then be used as a controllable parameter to tune the tunneling properties.

We show the conductance as a function of the energy $E$ in Fig.~\ref{fig6} for different gap values  $\Delta$ with $\phi=\pi$, $r_{0}=10\ \text{\AA{}}$, $N=200$ and $d=16\ \text{\AA{}}$ (top panel), $d=100\ \text{\AA{}}$ (bottom panel). 
Our results show that the behavior of the conductance in the form of peaks disappears along the energy line and stabilizes.
For $\Delta=0$ in Fig.~\ref{subfigure6a}, the conductance reaches the value null for $E=0$, with the appearance of other peaks which disappear. We observe that only the transmission of the same spin dominates the conductance. 
  Increasing $\Delta$ induces a displacement of the conductance at the level of the origin, as well as the emergence of other peaks, as seen in
Fig.~(\ref{subfigure6b},  \ref{subfigure6c}).
In contrast to the case where $\Delta =0$, the cancellation of the conductance is stronger at a certain value of $E\ne 0$, which is due to the contribution of spin inversion transmissions. 
The same holds true for $d=100\ \text{\AA{}}$, as shown in
Fig.~(\ref{subfigure6d}, \ref{subfigure6e}, \ref{subfigure6f}), except that the number of peaks multiplies and the interval between them becomes narrow.
As a result, we conclude that the gap $\Delta$ has a significant impact on the conductance of the rippled graphene superlattice.

	\section{Conclusion}
	
	The transmission and conduction of rippled graphene superlatice with a gap were studied. Our system was made of $N$ cells, each of which contained a ripple and a flat sheet. In our scenario, a ripple is a curved graphene surface with a radius of $r_ 0 $ and a mass term of $\Delta$ in the shape of an arc of a circle. 
	 The energy bands are obtained   by solving the Dirac equation in each region. As a result,  we demonstrated that adding a band gap $\Delta$ to our system appears to impact electron dispersion.
	 To calculate all transmissions and reflection channels, the energy spectrum is used with the transfer matrix method. 
	 
%
%
Our numerical results reveal and confirm that the discovery of electron transmission with opposing spin polarization is a crucial effect of $\Delta$. This was not the case in previous $\Delta=0$ investigations  \cite{puldak2020,smotlacha}. 
	Furthermore, an increase in $\Delta$ causes a decrease and suppression of the transmissions $T_{\uparrow}^{\uparrow} $ and $ T_{\downarrow}^{\downarrow} $, with the appearance of several peaks where the suppression is stronger. 	
 The transmission behavior shifts clearly at the origin of the energy $E=0$ as well as the interval between peaks. On the other hand, the spin inversion transmissions of electrons increased and suppressed with a higher number of peaks.
 The distance between peaks is narrowing, and peaks are also appearing where transmission is at its highest.
 Our findings reveal that the conductance is displaced at the origin level as well as the formation of other peaks. 
 Most notably, the add of a band gap  allows transmission and conductance to be controlled. 
 Finally, our results can be reproduced experimentally
 based on the experiment \cite{kuemmeth} proving that electron spin and orbital motion are related in nanotubes, which
may provide a solution for systems that meet technological requirements. 

	\end{document}